\documentclass{aa}

\usepackage{graphicx}
\usepackage{txfonts}
\usepackage{graphicx}
\usepackage{amsmath}
\usepackage[version=4]{mhchem}
\usepackage{siunitx}
\usepackage{pdflscape}
\usepackage{longtable,tabularx}
\usepackage{float}
\usepackage{multirow}
\usepackage{multicol}
\usepackage{url}
\usepackage{comment}
\usepackage{xcolor}
\usepackage{adjustbox}

\begin{document}

\title{A new meteor shower from comet 46P/Wirtanen expected in December 2023}

\titlerunning{46P new meteor shower}

\author{J. Vaubaillon \inst{1} \and
           Q.-Z. Ye       \inst{2,3}  \and
           A. Egal \inst{1,4,5} \and
           M. Sato \inst{6} \and
           D. E. Moser \inst{7}
          }

\institute{
IMCCE, CNRS, Observatoire de Paris, PSL Universit\'e, Sorbonne Universit\'e, Universit\'e de Lille 1, UMR 8028 du CNRS, 77 av. Denfert-Rochereau 75014 Paris, France  \quad \email{vaubaill@imcce.fr}
\and
 Department of Astronomy, University of Maryland, College Park, MD 20742, USA
\and
 Center for Space Physics, Boston University, Boston, MA 02215, USA
\and
Planétarium de Montréal, Espace pour la Vie, 4801 av. Pierre-de Coubertin, Montréal, Québec, Canada
\and
 Department of Physics and Astronomy, University of Western Ontario, London, Ontario, N6A 3K7, Canada
\and
National Astronomical Observatory of Japan, Tokyo, Japan
\and
NASA Meteoroid Environment Office, Marshall Space Flight Center, Huntsville, AL 35812, USA
}

\date{Received xx 2023; accepted XXX 2023}

  \abstract
   {Comet 46P/Wirtanen is a near-Earth object (NEO) for which no associated meteor shower has ever been reported.}
   {This study is aimed at improving our understanding of why there has been no observed shower activity for this NEO to date, as well as to consider whether any past activity could be uncovered from the post-prediction results.}
   {The usual dynamic tools for  meteoroid streams were used to describe the behavior of the particles ejected by the comet. The resulting modeled meteoroid stream was thoroughly inspected for collisions between the stream and the Earth.}
   {The results show a possible encounter forecast for December 12, 2023, between 8:00 and 12:30 UT. The slow entry velocity is typically known to cause dim meteors. The activity level of the shower is highly uncertain due to the absence of reported past showers.}
   {Overall, the most optimal observations on the forecasted day would be achieved from Eastern Australia, New Zealand, and Oceania. These observations will help constrain the size distribution of meteoroids from comet 46P/Wirtanen in the millimeter range.}


\maketitle

\section{Introduction}\label{sec:intro}

Comet 46P/Wirtanen was discovered in January 1948 by C. Wirtanen at Lick Observatory.
It is classified as a Jupiter-family comet (JFC), with the size of its nucleus estimated to be $1.2 - 1.4\; km$ by \cite{Lamy2004,Moulane2023} and $1.4 \pm 0.1\; km $ by \cite[][E. Howell, private comm]{Lejoly2022}.
Comet 46P was the initial target of the Rosetta mission \citep{Rickman1998}, until the probe launch delay led to the focus to be shifted onto comet 67P/Churyumov-Gerasimenko instead.
Its orbit causes relatively close encounters with Jupiter, with the latest taking place in 1972 at 0.276 $au$ from the giant planet.
Another very close encounter with the Earth in 2018 ($0.077\; au$) allowed for detailed characterizations to be performed. These revealed the comet is a hyper-active comet \citep{Moulane2023}, meaning that its active surface area is higher than expected.

46P is a near-Earth comet and, as such, it may potentially be the parent body of a meteor shower \citep{YeJenniskens2022}.
Moreover, an optical trail was detected by \cite{Farnham2019}.
\cite{MaslovMuzyko2017} did look for possible showers from the modelling of the associated meteoroid stream generation and subsequent evolution, but no confirmation was performed by significant observation.
The goal of this paper is to re-investigate the possibility of a meteor shower caused by 46P, based on several different models and approaches.
We find an encounter between the Earth and the stream in December 2023.
Section \ref{sec:meth} describes the method used.
Section \ref{sec:res} provides the results of the dynamics of the stream and the forecast shower in 2023.

\section{Method}\label{sec:meth}

The methods used are based on similar previous studies of the dynamics of meteoroid streams and its application to the forecast of meteor showers \citep{YeVaubaillon2021,YeVaubaillon2022}. In this case, several different models are considered. For each model, simulated meteoroids were ejected from 46P/Wirtanen at each return of the comet and integrated until the present epoch to analyze the current dust distribution near the Earth. Details on the simulation parameters used are presented in Table~\ref{tab:model_parameters}.

One model used was developed by \cite{Ye2016}; this model tested a classical pure-ice model \citep{Whipple1950} with ejection speeds multiplied by $0.1\times$, $1\times,$ and $5\times$ in order to cover different activity levels of the parent comet. For each scenario, about 1 million particles were generated and simulated.

Another model is based on \cite{Vaubaillon2005a,Vaubaillon2005b}. The comet perihelion passages taken into account here are restricted to the observed ones, i.e. since 1954. The dust production rate was calculated using the measured $Af\rho=150$~cm, where $Af\rho$ is a proxy of dust production \citep{AHearn1984}. A total of $650E+03$ particles were simulated for this model.

Additional simulations were conducted based on the model of \cite{Egal2019}. Several simulations were performed to investigate the influence of the diameter of the nucleus and the fraction of active area (from 20\% to 40\%). In conjunction with these independent simulations we tested two different dust size distributions for the comet, characterized by the differential size indices of $\alpha_\mathrm{d}=3.3,$ as found by \citet{Protopapa2021}, which is in line with typical comets \citep{Fulle2004} and the unusually steep $\alpha_\mathrm{d}=5.9$ as found by \citet{Kareta2023}. 

The model from \cite{Moser2004EMP,Moser2008EMP} was used in a fourth simulation, taking into account perihelion passages since 1900 and utilizing the meteoroid ejection model of \cite{Jones1996ASPC137}.  The active area fraction was set at 50\%. $57E+06$ particles were simulated for this model.

Lastly, the model by \cite{Sato2005} was also used to assess the possibility of a meteor shower from 46P in 2023.

\begin{table*}[!ht]
    \centering
    \begin{tabular}{clllll}
        \hline
        \hline
        Model & Ejection model & Perihelion passages &  $f_a$  & D (km) & Description \\
        \hline
        QY1 & 0.1$\times$W50  &  1900 - 2000  & & & \cite{Ye2016} \\
        QY2 & 1$\times$W50  & 1900 - 2000  & &  & \\
        QY3 & 5$\times$W50  & 1900 - 2000  & & & \\
        \hline
        JV &  CR97 & 1901 -  &  20\% & 1.2 & \cite{Vaubaillon2005a,Vaubaillon2005b}\\
        \hline
        AE1 &  CR97 & 1830 - 2050  &  20\% & 1.2 & \cite{Egal2019}\\
        AE2 &  CR97 & 1830 - 2050  & 40\% & 1.4 &  \\
        \hline
        DM &  JB96 & 1900 - 2050  &  50\% & 1.2 & \cite{Moser2004EMP,Moser2008EMP}\\
        \hline
        SW & fit & 1974   & & & \cite{Sato2005} \\
        \hline
        \hline
    \end{tabular}
    \caption{Physical characteristics of the 46P nucleus (diameter $D$ and fraction of active area $f_a$) and meteoroid ejection parameters considered for the simulations. (The QY models do not use this information.) Meteoroids were ejected from the nucleus within the limiting heliocentric distance $r_h \leq $ 3 AU, with speeds following the model of CR97 \citep{Crifo1997}, JB96 \citep{Jones1996ASPC137}, or W50 \citep{Whipple1950}. The model by \cite{Sato2005} explores the ejection velocity needed to bring the meteoroids to Earth for a given year, so the ejection velocity is tuned for a specific prediction.}
    \label{tab:model_parameters}
\end{table*}

\section{Results}\label{sec:res}

\subsection{Dynamics of the comet}\label{sec:dyncom}

\begin{figure}[!htbp]
\begin{center}
\includegraphics[width=0.45\textwidth,keepaspectratio]{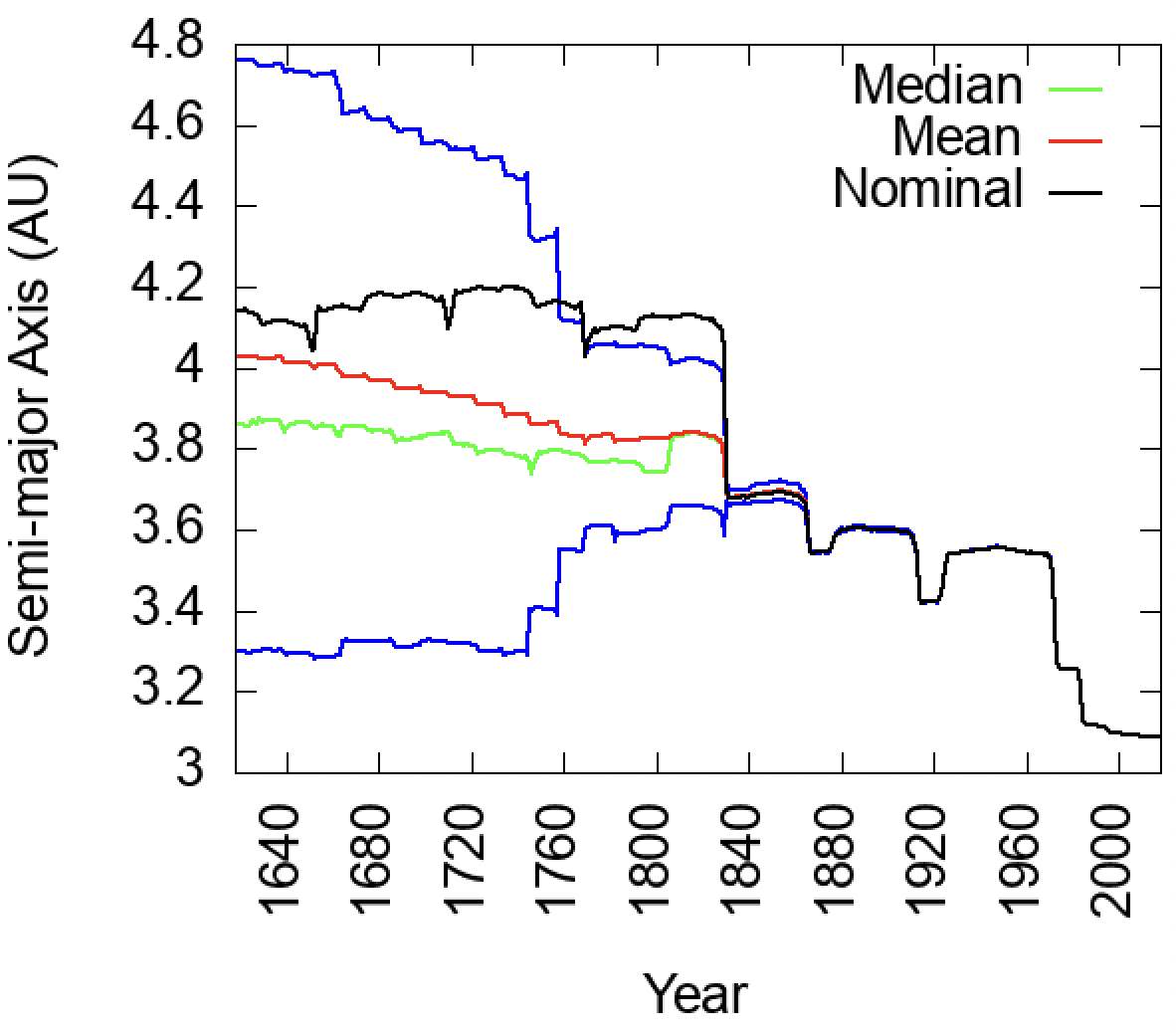}
\caption{Time-evolution of the nominal (black), average (red) and median (green) orbital elements of 1000 clones of comet 46P/Wirtanen generated in 2018. The blue curve represents the one standard deviation range about the average orbital element. All comet clones follow a similar evolution until 1830.}
\label{fig:comdyn}
\end{center}
\end{figure}

To determine the reliability of 46P ephemeris, we simulated the past orbital evolution of a thousand clones of the comet's orbit. We used as starting conditions the orbital solution K181/24 provided by the JPL and examined the dispersion of the clones' osculating elements. Since the clones displayed a similar dispersion behavior for each orbital element, we only illustrate in Figure \ref{fig:comdyn} the backward evolution of the comets' semi-major axis.

We see that multiple close encounters with Jupiter (illustrated by sudden jumps in semi-major axis) significantly modified 46P's orbit in the past. However, the clones' dispersion remains small until 1830, indicating that the comet can be confidently traced back for about two centuries. All the simulations presented in Table \ref{tab:model_parameters} were thus restricted to perihelion passages posterior to 1830, when the ephemerides of the comet are well-determined.


\subsection{Past apparitions }\label{sec:past}

Like many JFC meteoroid streams, the dynamics shows a rather complicated behavior, with the reversal process \citep{Vaubaillon2004} at play.
Perturbed meteoroid trails were found to intersect the Earth's orbit in the past, although the spatial density of the dust is not easy to constrain.

Unlike \cite{MaslovMuzyko2017}, no significant encounter was found in 2012 or 2017 in the QY and JV simulations. In the AE and DM models, thin particle trails formed by particles ejected from 46P between 1925 and 1950 were found to intersect the Earth's orbit in 2012 and 2017. However, the spatial density of the trails makes any past detection of the shower unlikely. Differences between the models can be explained by different consideration of the non-gravitational forces affecting the comet's and meteoroids' motion, as well as the ejection conditions. In contrast, all models were found to produce some encounters with the Earth in 2007 and 2018, mainly caused by the 1974 trail, which may have been recorded by dedicated meteor detection networks.

The orbit of the comet causes the geocentric velocity of the associated meteoroids to be "low," namely: $V_g \sim 10.2 \; km.s^{-1}$ (see also table \ref{tab:pred}).
Due to the low geocentric velocity of the meteoroids, we expect only mm-sized meteoroids to be detectable with video networks, while particles smaller than a millimeter may be observed with radio instruments. A similar situation occurred in 2021, with the first apparition of the Arid meteor shower. Despite the low geocentric velocity of the meteoroids ($V_g \le$ 16 km/s), the Arids was detected by the SAAMER \citep[Southern Argentina Agile Meteor Radar-Orbital System][]{Janches2023}. Although the meteor radiants were located at low declinations in the Southern hemisphere, multiple Arids were even recorded by the Canadian Meteor Orbit Radar \citep[CMOR][]{Brown2008,Brown2010} because of the gravitational bending of their trajectory close to the Earth.

Encouraged by the Arids detection, we searched the CMOR database for possible apparitions of the shower produced by 46P in 2007, 2012, 2017, and 2018. We performed a wavelet analysis of the radar observations based on the modelled characteristics of the shower, as described in \cite{Brown2010,Egal2022}. Unfortunately, we found no conclusive detection of the shower in past CMOR observations. To our knowledge, no observation of the shower has ever been reported in the literature.

\subsection{Predictions for December 2023}\label{sec:pred}

\begin{figure*}[!htbp]
\begin{center}
\begin{tabular}{cc}
\includegraphics[width=0.4\textwidth,keepaspectratio]{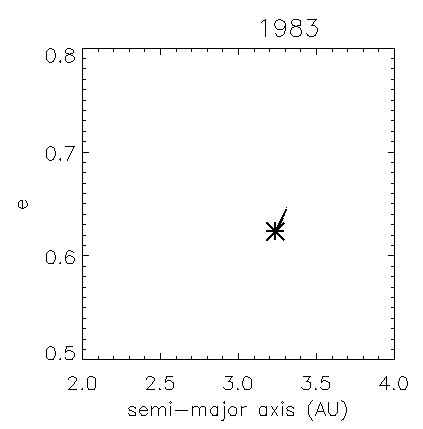}& 
\includegraphics[width=0.4\textwidth,keepaspectratio]{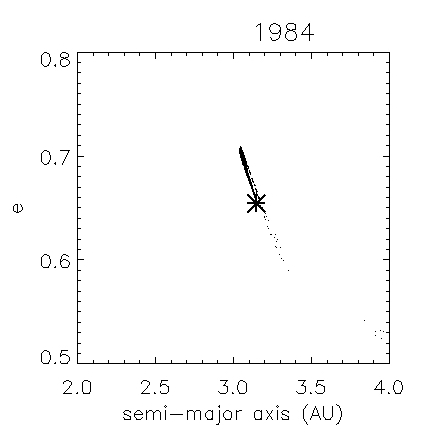} \\
\end{tabular}
\caption{Dynamics of the 1974 trail: Semi-major axis vs eccentricity before (left) and after (right) the close encounter with Jupiter. The star represents the position of comet 46P.}
\label{fig:dyn}
\end{center}
\end{figure*}

All the independent models presented in Section \ref{sec:meth} predict an encounter between the Earth and the trail ejected in 1974. This trail shows a rather complicated dynamical evolution; in 1984, a close encounter of the stream with Jupiter caused a dramatic change in the orbital elements of the meteoroids. Figure \ref{fig:dyn} highlights such change by comparing the orbital elements before and after the encounter. The direct consequence is the leading position of the trail with respect to the comet, in particular in 2023 (ibid).

\begin{figure}[!htbp]
\begin{center}
\includegraphics[trim={0.2cm 0.1cm 0.1cm 0.2cm}, clip, width=0.45\textwidth,keepaspectratio]{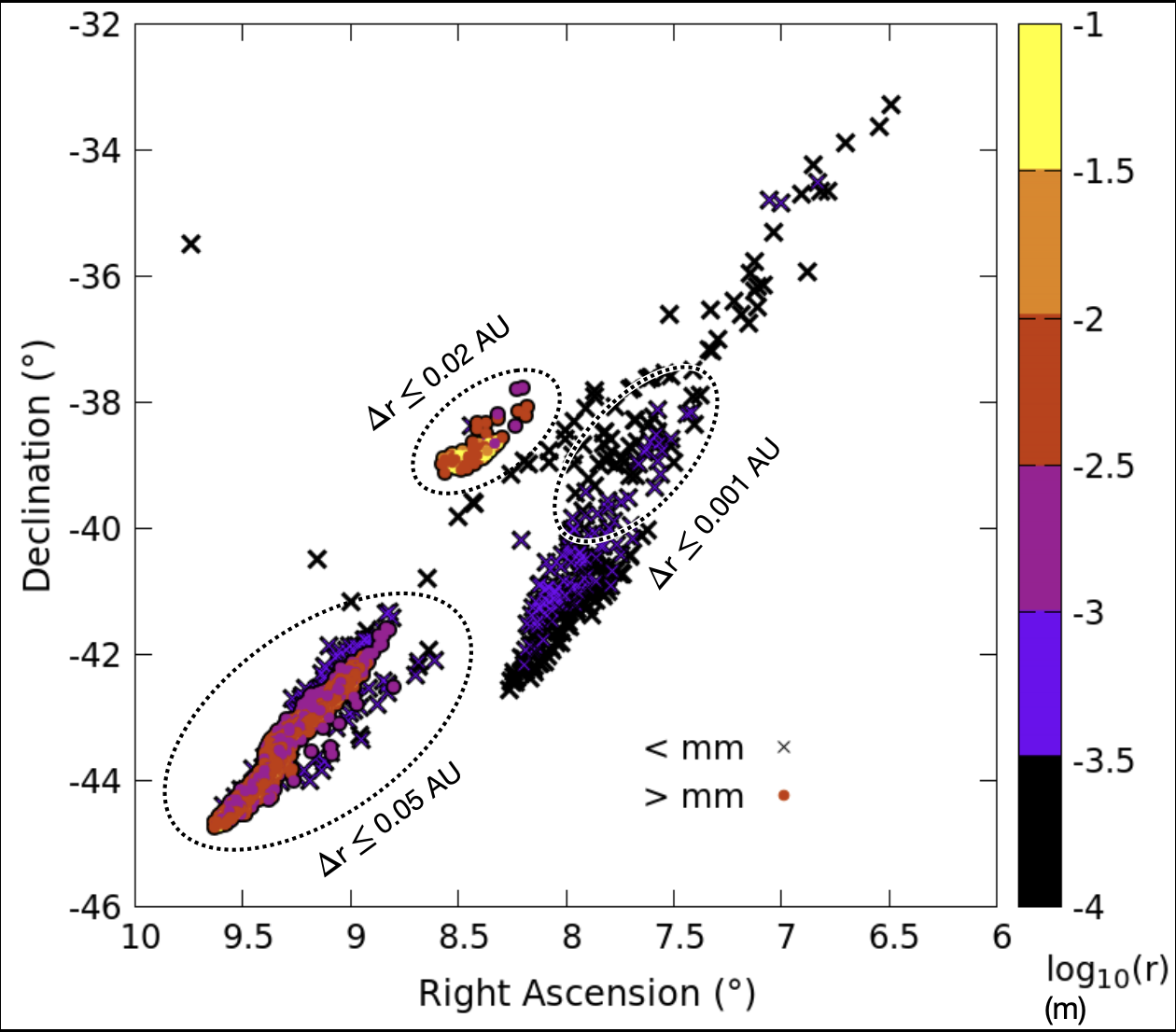}
\caption{Geocentric radiant of the simulated meteoroids in 2023 (AE1), color-coded as function of the particle's radius. Sub-mm particles are represented by crosses, while mm-sized particles that should be visible to radio and video instruments are represented by filled circles. Particles approaching Earth below $\Delta$r (AU) are retained as potential impactors. Depending on the $\Delta$r threshold considered, several clusters of radiants may be detectable in December 2023. However, only $\le$ mm-sized particles intersect the Earth's orbit in the AE model. }
\label{fig:SFD}
\end{center}
\end{figure}

The expected meteoroid spatial density is very low, although it is not easy to constrain, and highly depends on the considered model.
In particular, a relatively high ejection velocity (QY3)~brings more particles to the Earth than a regular ejection process. The size distribution of the meteoroids released at usual ejection velocities suggest that only the smallest particles ($\le 1$ mm) intersect the Earth's orbit in 2023, while meteoroids up to a few centimeters in radius approach the Earth within 0.05 AU (cf. figure \ref{fig:SFD}).

\begin{figure*}[!htbp]
\begin{center}
\begin{tabular}{cc}
\includegraphics[width=0.45\textwidth,keepaspectratio]{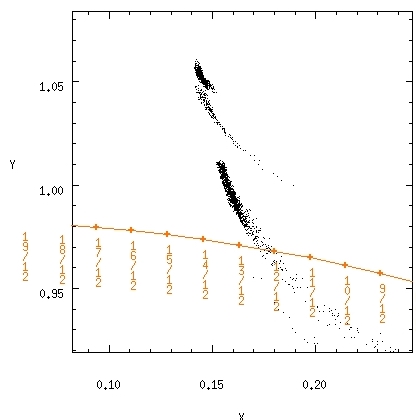}
\includegraphics[width=0.45\textwidth,keepaspectratio]{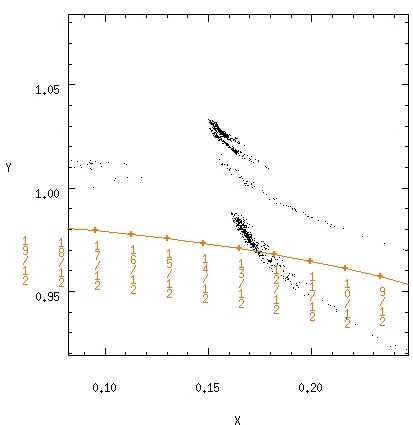}
\end{tabular}
\caption{Node of the meteoroids compared to the path of the Earth in 2007 (left) and 2023 (right) from the JV model.}
\label{fig:node}
\end{center}
\end{figure*}

A summary of the predictions is provided in table \ref{tab:pred}. The timing of the shower differs from one model to another, depending on the ejection model and the timing and distance criteria adopted to select meteor-producing particles. Most models involving the simulation of both small and large particles locate the activity on December 12, between 8h and 12h30 UT (JV, QY, AE $\le$ mm, DM). Models retaining only the contribution of large particles (related to the core of the trail, which does not intersect the Earth's orbit - cf. Fig. \ref{fig:node}) predict the maximum activity to occur later that day or early on December 13 (e.g., Sato, AE > mm). Given the uncertainties on the stream's evolution, we find the different models to be in good agreement with each other, and to predict clear meteor activity on December 12, 2023.

\begin{table*}
    \centering
    \caption{Predictions of the 2023 shower from different models. $\Delta_0$: distance between the trail and the Earth at the time of maximum. Symbols: $V_{geo}$ is the geocentric velocity; $\alpha$ and $\delta$ are the location of the radiant; and $ZHR$ is the expected level of the shower.}
    \label{tab:pred}
     \begin{adjustbox}{width=\textwidth}
    \begin{tabular}{lccclcccc}
    \hline
    \hline
    Model   & Trail & $\Delta_0$    & Time of max   & $V_{geo}$     & $\alpha$  & $\delta$  & $ZHR$ \\
            &       &   [$au$]         &   [UT]        & [$km.s^{-1}$] & [$\deg$]  & [$\deg$]  & [$hr^{-1}$] \\
    \hline
    JV & 1974 & 0.00518 & 2023-12-12T10:15 & $10.2$    & $7.3$   & $-38.5$ & 5-50 (?) \\
    \hline
    QY1 -- $0.1\times$ Whipple & 1974 & 0.00008 & 2023-12-12T08:23$\pm0.6$~hr & $10.28\pm0.03$    & $7.5^\circ \pm0.1^\circ$   & $-38.1^\circ \pm0.2^\circ$ & 2--3 \\
    QY2 -- $1\times$ Whipple & 1974 & 0.00010 & 2023-12-12T09:32$\pm0.5$~hr & $10.28\pm0.03$    & $7.5^\circ \pm0.1^\circ$   & $-38.1^\circ \pm0.3^\circ$ & 10--20 \\
    QY3 -- $5\times$ Whipple & 1974 & 0.00013 & 2023-12-12T09:44$\pm0.6$~hr & $10.30\pm0.01$    & $7.6^\circ \pm0.1^\circ$   & $-38.6^\circ \pm0.4^\circ$ & 10--15 \\
    \hline
    \cite{Sato2005} &  1974  & 0.00645  & 2023-12-12T20:06  & 10.3  & 7.9  & -40.4  &  5  \\
    \hline
    DM  &  1974  & 0.003  & 2023-12-12T12:25  & 10.3  & 7.8  & $-39.8$  &  5  \\ 
    \hline
    AE1 $<$ mm, $\Delta r$ < 0.02 AU & 1974 & <0.001 & 2023-12-12T11:34 $\pm$ 0.5h & 9.8 $\pm$ 0.02 & 8.4 $\pm$ 0.3 & -38.8 $\pm$ 0.2 & <10 \\
    AE2 $<$ mm, $\Delta r$ < 0.02 AU & 1974 & <0.001 & 2023-12-12T09:52 $\pm$ 0.5h &  10.2 $\pm$ 0.01 &  7.6 $\pm$ 0.1 & -38.0 $\pm$ 0.2 & <10 \\
    \hline
    AE1 $>$ mm, $\Delta r$ < 0.02 AU & 1900-1908+ & <0.05 & 2023-12-12T17:05 $\pm$ 0.7h & 9.8 $\pm$ 0.02 & 8.4 $\pm$ 0.3 & -38.8 $\pm$ 0.2 & <10 \\
    AE1 $>$ mm, $\Delta r$ < 0.05 AU & 1900-1908+ & <0.05 & 2023-12-13T06:26 $\pm$ 3h & 10.0 $\pm$ 0.3 & 9.19 $\pm$ 0.3 & -43.03 $\pm$ 1.5 & <10 \\
    AE2 $>$ mm, $\Delta r$ < 0.02 AU & 1900-1945 & <0.05 & 2023-12-13T01:01 $\pm$ 1h & 9.9 $\pm$ 0.04 & 9.0 $\pm$ 0.2 & -41.9 $\pm$ 1.2 & <15 \\
    \hline
    \hline
    \end{tabular}
    \end{adjustbox}
\end{table*}

\begin{figure*}[!htbp]
\begin{center}
\begin{tabular}{cc}
\includegraphics[width=0.4\textwidth,keepaspectratio]{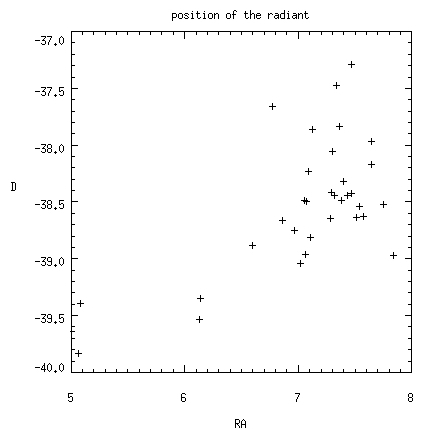}
\includegraphics[width=0.4\textwidth,keepaspectratio]{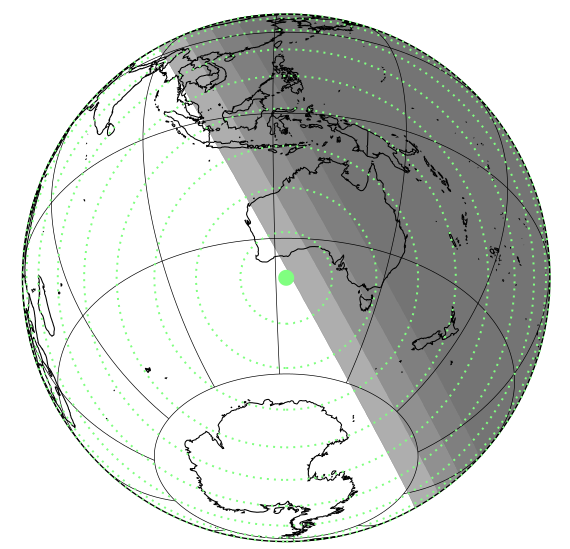}
\end{tabular}
\caption{Location of expected shower radiant (units are $\deg$) from the JV model, shown in the left panel. Shower visibility is given on the right.}
\label{fig:radiant}
\end{center}
\end{figure*}

Similarly to the new Arid meteor shower caused by comet 15P/Finlay in 2021 \citep{Vaubaillon2020}, the level of the shower is hard to grasp, since no observed past outburst can help calibrate the forecast. Because of the low encounter velocity ($10.2 \; km.s^{-1}$), the shower produced by 46P is expected to be rich in dim meteors, with only the largest meteoroids (if any) being detectable by optical detectors. In addition, we expect quite a spread in the modelled radiant location (see Fig. \ref{fig:radiant}); and it is worth recalling that the observed radiant dispersion is usually greater than the forecast dispersion.

On the other hand, the dynamical history of comet 46P may argue in favor of an enhanced meteoroid production in the past. \cite{Moulane2023} found the nucleus to be hyperactive, meaning that an equivalent of $40\%$ of its surface is active.
This is a huge value compared to other comets and may indicate an elevated meteoroid production rate, assuming the dust production rate is proportional to the water production rate, as reported in the study cited here.
In addition, the dynamical analysis in Section \ref{sec:dyncom} shows that the perihelion distance was larger in the past (at least for the time period considered here). This argues in favor of a quite high activity during the 1974 passage, namely, right after the perihelion distance decreased following the close encounter with Jupiter in 1972 \citep[similarly to comet 240P/NEAT, see][]{Kelley2019}.  However, the absence of detection of any trail using an IR telescope \citep{YeJenniskens2022} is puzzling, since a high production rate would create a dense meteoroid stream that may be detected.

Attempts to reproduce the possible high activity of the comet in the past were performed by modifying the fraction of active area, $f_a$, (from 20\% to 40\%) in the simulations. Different $f_a$ values were found to have a significant influence on the predicted time, but only a moderate impact on the meteor rates (cf. table \ref{tab:pred}). 

A more challenging parameter is the meteoroids size frequency distribution at meteor-producing meteoroid sizes that can only be guessed (by e.g., considering a constant index from $\mu$-sized particles detected in visible wavelengths in the comet coma). \cite{Kareta2023} found a very steep differential index of $\alpha_d>5.9$ for sizes in the range $[0.5;50\; \mu m]$.
In contrast, \cite{Protopapa2021} found a more usual value of $\alpha_d \sim 3.3$.
All things considered, the expected $ZHR$ in Table \ref{tab:pred} should be seen more like very rough estimates rather than accurate values.

Despite the uncertainties  on the activity level of the comet since 1900, all the predictions of in Table \ref{tab:pred} point toward a noticeable meteor activity on December 12, 2023.
The shower will be best visible from North and West Australia, Papua New Guinea, New Zealand, and Indonesia (see Fig. \ref{fig:radiant}).

\section{Conclusion}\label{sec:conclu}

After examining the dynamics of comet 46P/Wirtanen and its associated meteoroid stream, we have shown that the orbit of the comet is stable enough for the time span considered here.
Several past encounters with the Earth and the stream were found, but no observations were reported to our knowledge.
This is probably due to the quite unusual ejection velocity needed to bring large particles to the Earth.
We predict the birth of a new shower on December 12,  2023, between 8:00 and 12:30 UT.
The exact time of the maximum depends on the considered size frequency distribution and may vary by a few hours.
The location of the radiant is close to the $\lambda$-Sculptoris star, so a possible name for the shower is $\lambda$-Sculptorids.
The observation and subsequent report of this shower would be of tremendous scientific interest since it would put constraints on the size frequency distribution of large particles for the comet.
However, observations will be challenging due to the low entry velocity and the relatively small sizes of the meteoroids.
Nevertheless, we strongly encourage meteor enthusiasts to perform scientific observations and send their reports to the International Meteor Organization (IMO).

\section*{Acknowledgement}\label{sec:ack}

We are thankful to Josselin Desmars (IMCCE, Paris Observatory, PSL) for providing Fig. \ref{fig:radiant} and to Peter Brown (U.W.O, Canada) for digging in the CMOR data. QY was supported by NASA grant 80NSSC21K0156, and AE was supported in part by the NASA Meteoroid Environment Office under cooperative agreement 80NSSC21M0073.

\bibliographystyle{aa} 
\bibliography{main}

\begin{thebibliography}{30}
\expandafter\ifx\csname natexlab\endcsname\relax\def\natexlab#1{#1}\fi

\bibitem[{{A'Hearn} {et~al.}(1984){A'Hearn}, {Schleicher}, {Millis}, {Feldman}, \& {Thompson}}]{AHearn1984}
{A'Hearn}, M.~F., {Schleicher}, D.~G., {Millis}, R.~L., {Feldman}, P.~D., \& {Thompson}, D.~T. 1984, \aj, 89, 579

\bibitem[{{Brown} {et~al.}(2008){Brown}, {Weryk}, {Wong}, \& {Jones}}]{Brown2008}
{Brown}, P., {Weryk}, R.~J., {Wong}, D.~K., \& {Jones}, J. 2008, Earth Moon and Planets, 102, 209

\bibitem[{{Brown} {et~al.}(2010){Brown}, {Wong}, {Weryk}, \& {Wiegert}}]{Brown2010}
{Brown}, P., {Wong}, D.~K., {Weryk}, R.~J., \& {Wiegert}, P. 2010, \icarus, 207, 66

\bibitem[{{Crifo} \& {Rodionov}(1997)}]{Crifo1997}
{Crifo}, J.~F. \& {Rodionov}, A.~V. 1997, Icarus, 127, 319

\bibitem[{{Egal} {et~al.}(2022){Egal}, {Brown}, {Wiegert}, \& {Kipreos}}]{Egal2022}
{Egal}, A., {Brown}, P.~G., {Wiegert}, P., \& {Kipreos}, Y. 2022, \mnras, 512, 2318

\bibitem[{{Egal} {et~al.}(2019){Egal}, {Wiegert}, {Brown}, {Moser}, {Campbell-Brown}, {Moorhead}, {Ehlert}, \& {Moticska}}]{Egal2019}
{Egal}, A., {Wiegert}, P., {Brown}, P.~G., {et~al.} 2019, \icarus, 330, 123

\bibitem[{{Farnham} {et~al.}(2019){Farnham}, {Kelley}, {Knight}, \& {Feaga}}]{Farnham2019}
{Farnham}, T.~L., {Kelley}, M. S.~P., {Knight}, M.~M., \& {Feaga}, L.~M. 2019, \apjl, 886, L24

\bibitem[{{Fulle} {et~al.}(2004){Fulle}, {Barbieri}, {Cremonese}, {Rauer}, {Weiler}, {Milani}, \& {Ligustri}}]{Fulle2004}
{Fulle}, M., {Barbieri}, C., {Cremonese}, G., {et~al.} 2004, \aap, 422, 357

\bibitem[{{Janches} {et~al.}(2023){Janches}, {Bruzzone}, {Weryk}, {Hormaechea}, \& {Brunini}}]{Janches2023}
{Janches}, D., {Bruzzone}, J.~S., {Weryk}, R.~J., {Hormaechea}, J.~L., \& {Brunini}, C. 2023, \psj, 4, 165

\bibitem[{{Jones} \& {Brown}(1996)}]{Jones1996ASPC137}
{Jones}, J. \& {Brown}, P. 1996, in Astronomical Society of the Pacific Conference Series, Vol. 104, IAU Colloq. 150: Physics, Chemistry, and Dynamics of Interplanetary Dust, ed. B.~A.~S. {Gustafson} \& M.~S. {Hanner}, 137

\bibitem[{{Kareta} {et~al.}(2023){Kareta}, {Noonan}, {Harris}, \& {Springmann}}]{Kareta2023}
{Kareta}, T., {Noonan}, J.~W., {Harris}, W.~M., \& {Springmann}, A. 2023, \psj, 4, 85

\bibitem[{{Kelley} {et~al.}(2019){Kelley}, {Bodewits}, {Ye}, {Farnham}, {Bellm}, {Dekany}, {Duev}, {Helou}, {Kupfer}, {Laher}, {Masci}, {Prince}, {Rusholme}, {Shupe}, {Soumagnac}, \& {Zolkower}}]{Kelley2019}
{Kelley}, M. S.~P., {Bodewits}, D., {Ye}, Q., {et~al.} 2019, \apjl, 886, L16

\bibitem[{{Lamy} {et~al.}(2004){Lamy}, {Toth}, {Fernandez}, \& {Weaver}}]{Lamy2004}
{Lamy}, P.~L., {Toth}, I., {Fernandez}, Y.~R., \& {Weaver}, H.~A. 2004, in Comets II, ed. M.~C. {Festou}, H.~U. {Keller}, \& H.~A. {Weaver}, 223

\bibitem[{{Lejoly} {et~al.}(2022){Lejoly}, {Harris}, {Samarasinha}, {Mueller}, {Howell}, {Bodnarik}, {Springmann}, {Kareta}, {Sharkey}, {Noonan}, {Bedin}, {Bosch}, {Brosio}, {Bryssinck}, {de Vanssay}, {Hambsch}, {Ivanova}, {Krushinsky}, {Lin}, {Manzini}, {Maury}, {Moriya}, {Ochner}, \& {Oldani}}]{Lejoly2022}
{Lejoly}, C., {Harris}, W., {Samarasinha}, N., {et~al.} 2022, \psj, 3, 17

\bibitem[{{Maslov} \& {Muzyko}(2017)}]{MaslovMuzyko2017}
{Maslov}, M.~P. \& {Muzyko}, E.~I. 2017, Earth Moon and Planets, 119, 85

\bibitem[{{Moser} \& {Cooke}(2004)}]{Moser2004EMP}
{Moser}, D.~E. \& {Cooke}, W.~J. 2004, Earth Moon and Planets, 95, 141

\bibitem[{{Moser} \& {Cooke}(2008)}]{Moser2008EMP}
{Moser}, D.~E. \& {Cooke}, W.~J. 2008, Earth Moon and Planets, 102, 285

\bibitem[{{Moulane} {et~al.}(2023){Moulane}, {Jehin}, {Manfroid}, {Hutsem{\'e}kers}, {Opitom}, {Shinnaka}, {Bodewits}, {Benkhaldoun}, {Jabiri}, {Hmiddouch}, {Vander Donckt}, {Pozuelos}, \& {Yang}}]{Moulane2023}
{Moulane}, Y., {Jehin}, E., {Manfroid}, J., {et~al.} 2023, \aap, 670, A159

\bibitem[{{Protopapa} {et~al.}(2021){Protopapa}, {Kelley}, {Woodward}, \& {Yang}}]{Protopapa2021}
{Protopapa}, S., {Kelley}, M. S.~P., {Woodward}, C.~E., \& {Yang}, B. 2021, \psj, 2, 176

\bibitem[{{Rickman} \& {Jorda}(1998)}]{Rickman1998}
{Rickman}, H. \& {Jorda}, L. 1998, Advances in Space Research, 21, 1491

\bibitem[{{Vaubaillon} {et~al.}(2005{\natexlab{a}}){Vaubaillon}, {Colas}, \& {Jorda}}]{Vaubaillon2005a}
{Vaubaillon}, J., {Colas}, F., \& {Jorda}, L. 2005{\natexlab{a}}, \aap, 439, 751

\bibitem[{{Vaubaillon} {et~al.}(2005{\natexlab{b}}){Vaubaillon}, {Colas}, \& {Jorda}}]{Vaubaillon2005b}
{Vaubaillon}, J., {Colas}, F., \& {Jorda}, L. 2005{\natexlab{b}}, \aap, 439, 761

\bibitem[{{Vaubaillon} {et~al.}(2020){Vaubaillon}, {Egal}, {Desmars}, \& {Baillie}}]{Vaubaillon2020}
{Vaubaillon}, J., {Egal}, A., {Desmars}, J., \& {Baillie}, K. 2020, WGN, Journal of the International Meteor Organization, 48, 29

\bibitem[{{Vaubaillon} {et~al.}(2004){Vaubaillon}, {Lamy}, \& {Jorda}}]{Vaubaillon2004}
{Vaubaillon}, J., {Lamy}, P., \& {Jorda}, L. 2004, Earth Moon and Planets, 95, 75

\bibitem[{{Watanabe} {et~al.}(2005){Watanabe}, {Sato}, \& {Kasuga}}]{Sato2005}
{Watanabe}, J.-I., {Sato}, M., \& {Kasuga}, T. 2005, \pasj, 57, L45

\bibitem[{{Whipple}(1950)}]{Whipple1950}
{Whipple}, F.~L. 1950, \apj, 111, 375

\bibitem[{{Ye} \& {Jenniskens}(2022)}]{YeJenniskens2022}
{Ye}, Q. \& {Jenniskens}, P. 2022, arXiv e-prints, arXiv:2209.10654

\bibitem[{{Ye} \& {Vaubaillon}(2022)}]{YeVaubaillon2022}
{Ye}, Q. \& {Vaubaillon}, J. 2022, \mnras, 515, L45

\bibitem[{{Ye} {et~al.}(2021){Ye}, {Vaubaillon}, {Sato}, \& {Maslov}}]{YeVaubaillon2021}
{Ye}, Q., {Vaubaillon}, J., {Sato}, M., \& {Maslov}, M. 2021, The Astronomer's Telegram, 14947, 1

\bibitem[{{Ye} {et~al.}(2016){Ye}, {Hui}, {Brown}, {Campbell-Brown}, {Pokorn{\'y}}, {Wiegert}, \& {Gao}}]{Ye2016}
{Ye}, Q.-Z., {Hui}, M.-T., {Brown}, P.~G., {et~al.} 2016, \icarus, 264, 48

\end{thebibliography}

\end{document}